\documentstyle[aps,prd,psfig]{revtex}

\makeatletter
\newbox\slashbox \setbox\slashbox=\hbox{\large$/$}
\def\pslash#1{\setbox\@tempboxa=\hbox{$#1$}
  \@tempdima=0.5\wd\slashbox \advance\@tempdima 0.5\wd\@tempboxa
  \copy\slashbox \kern-\@tempdima \box\@tempboxa}
\def\slash{\protect\pslash}
\makeatother

\begin{document}

\draft

\title{The microscopic spectrum of the QCD Dirac operator with finite
       quark masses}

\author{T. Wilke$^1$, T. Guhr$^1$, and T. Wettig$^2$}

\address{%
  $^1$Max Planck Institut f\"ur Kernphysik, Postfach 103980, D-69029
  Heidelberg, Germany\\ 
  $^2$Institut f\"ur Theoretische Physik, Technische Universit\"at
  M\"unchen, D-85747 Garching, Germany}

\date{November 10, 1997}

\maketitle

\begin{abstract}
  We compute the microscopic spectrum of the QCD Dirac operator in the
  presence of dynamical fermions in the framework of random-matrix
  theory for the chiral Gaussian unitary ensemble.  We obtain results
  for the microscopic spectral correlators, the microscopic spectral
  density, and the distribution of the smallest eigenvalue for an
  arbitrary number of flavors, arbitrary quark masses, and arbitrary
  topological charge.
\end{abstract}


\section{Introduction}
\label{sec1}

The low-lying eigenvalues of the QCD Dirac operator are of great
importance for the understanding of the spontaneous breaking of chiral
symmetry in QCD.  In Euclidean space, the Dirac operator reads
$i\slash D=i\slash\partial+g(\lambda^a/2)\slash A^a$, where $g$ is the
coupling constant, $\lambda^a$ are the generators of SU($N$)-color,
and $A_\mu^a$ are the gauge fields.  The corresponding eigenvalue
equation is $i\slash D \psi_i=\lambda_i \psi_i$ with real eigenvalues
$\lambda_i$.  Because the Dirac operator anticommutes with $\gamma_5$,
the eigenvalues either occur in pairs $\pm\lambda_i$ or are zero.  Let
$\rho(\lambda) = \langle \sum_i\delta(\lambda-\lambda_i)\rangle_A$ be
the eigenvalue density of the Dirac operator averaged over gauge field
configurations.  The Banks-Casher formula,
$\langle\bar\psi\psi\rangle=\pi\rho(0)/V$ \cite{Bank80}, relates the
order parameter of chiral symmetry breaking,
$\langle\bar\psi\psi\rangle$, to the density of eigenvalues of the
Dirac operator near zero virtuality.  Here, it is essential that the
thermodynamic limit be taken before the chiral limit.  Note also that
$\rho(\lambda)$ is normalized to the space-time volume $V$.  In the
real world chiral symmetry is spontaneously broken and
$\langle\bar\psi\psi\rangle\ne 0$.  This implies that the low-lying
Dirac eigenvalues must be spaced like $\sim 1/V$.  Therefore, it is
natural to magnify this region of the spectrum by a factor of $V$.
This was first suggested by Shuryak and Verbaarschot \cite{Shur93} who
introduced the so-called microscopic spectral density defined by
\begin{equation}
  \label{eq1.1}
  \rho_s(z)=\lim_{V\to \infty}\frac{1}{V\Sigma}\;
  \rho\left(\frac{z}{V\Sigma}\right)\:,
\end{equation}
where $\Sigma$ is the absolute value of the chiral condensate.  Based
on an analysis of sum rules derived by Leutwyler and Smilga
\cite{Leut92} for the inverse powers of the Dirac eigenvalues in a
finite volume, Shuryak and Verbaarschot conjectured that the
microscopic spectral density should be a universal quantity which
depends only on global symmetries, the number of flavors $N_f$, and
the topological charge $\nu$.  What is the meaning of universality in
this statement?  In principle, one could compute $\rho_s$ directly
from full QCD.  In practice, of course, the complexity of the QCD
Lagrangian makes this impossible.  However, if $\rho_s$ is universal,
one can try to compute it in another theory which is simpler than QCD
but has the same symmetries.  The simplest theory in which such a
calculation is possible is random-matrix theory (RMT).  Whether the
results obtained in RMT correctly describe the microscopic spectrum of
full QCD is a question that can only be answered by empirical
evidence, most importantly by large-scale lattice QCD simulations.

Let us briefly recall how a chiral random-matrix model is constructed.
In a chiral basis in Euclidean space, the Dirac operator has a block
structure since it only couples states of opposite chirality.  In RMT,
the matrix of the Dirac operator is replaced by a random matrix $W$
with suitable symmetry properties,
\begin{equation}
  \label{eq1.2}
  \left[\matrix{im&i\slash{D}\cr (i\slash{D})^\dagger&im}\right]\to
  \left[\matrix{im&W\cr W^\dagger&im}\right]\:,
\end{equation}
where we have also added a quark mass $m$.  If $W$ has $N+\nu$ rows
and $N$ columns, then the Dirac matrix in (\ref{eq1.2}) (without the
mass term) has $N$ positive and $N$ negative eigenvalues,
respectively, and $\nu$ zero modes (we assume $\nu\ge 0$ and $\nu\ll
N$).  We can identify $\nu$ with the topological charge and
$2N+\nu\approx 2N$ with the volume $V$.  In full QCD with $N_f$
flavors, the weight function used in averaging over gauge field
configurations contains the gluonic action in the form $\exp(-S_{\rm
gl})$ and $N_f$ fermion determinants.  In the random-matrix model, the
gluonic part of the weight function is replaced by a convenient
distribution of the random matrix $W$, usually a Gaussian distribution
of the individual matrix elements (which are uncorrelated).  Since the
fermion determinants can be expressed as products over the Dirac
eigenvalues they can also be taken into account in the random-matrix
model.  The symmetries of $W$ are determined by the anti-unitary
symmetries of the Dirac operator.  They were classified by
Verbaarschot in Ref.~\cite{Verb94a}.  Depending on the number of
colors and the representation of the fermions the matrix elements of
$W$ are real, complex, or quaternion real.  The corresponding
random-matrix ensembles are called chiral Gaussian orthogonal (chGOE),
unitary (chGUE), and symplectic (chGSE) ensemble, respectively.  The
microscopic spectral density has been computed analytically in all
three ensembles for all $N_f$ and $\nu$ in the chiral limit
\cite{Verb93,Verb94d,Naga95}.

Evidence in support of the conjecture that $\rho_s$ is a universal
function has been accumulated in a number of recent studies which we
list here. The moments of $\rho_s$ generate the Leutwyler-Smilga sum
rules \cite{Verb93}.  The result for $\rho_s$ is insensitive to the
probability distribution of the random matrix \cite{Brez96,Nish96}.
Lattice data for the valence quark mass dependence of the chiral
condensate could be understood using the analytical expression for
$\rho_s$ \cite{Chan95,Verb96a}.  The functional form of $\rho_s$ does
not change at finite temperature \cite{Jack96b}.  The analytical
result for $\rho_s$ is found in the Hofstadter model for universal
conductance fluctuations \cite{Slev93}.  For an instanton liquid,
$\rho_s$ shows quite good agreement with the random-matrix result
\cite{Verb94b}.  Recently, a high-statistics lattice calculation
directly demonstrated agreement between the random-matrix results and
lattice data, not only for $\rho_s$ but also for the distribution of
the smallest eigenvalue and for the microscopic spectral two-point
correlator \cite{Berb97a}. (This study was done for gauge group SU(2)
with staggered fermions in the quenched approximation.)  We believe
that the universality of $\rho_s$ is now firmly established.

One of the most interesting and most important problems of present-day
lattice simulations is the inclusion of dynamical fermions.  This is
essential for a realistic description of hadronic properties.
Unfortunately, it is computationally expensive to include the fermion
determinants in a Monte-Carlo simulation.  Moreover, the computational
cost increases with lower quark mass, thus it is extremely difficult
to take the chiral limit on the lattice.  Any analytical insight into
the distribution of the Dirac eigenvalues in the presence of massive
dynamical quarks will therefore be helpful.  It is the purpose of this
work to extend the random-matrix results obtained in the chiral limit
to the general case of massive quarks to be able to compare with
actual lattice calculations.  This issue was first discussed in
Ref.~\cite{Jurk96}.  The most interesting case of QCD with 3 colors
and fermions in the fundamental representation corresponds to the
chGUE of RMT on which we concentrate here.  In what range of the quark
mass can we expect the random-matrix results to be in agreement with
full QCD?  Gasser and Leutwyler \cite{Gass87} have shown that in the
range $1/\Lambda\ll L\ll 1/m_\pi$ the dependence of the finite-volume
QCD partition function on the quark mass $m$ is determined solely by
global symmetries ($\Lambda$ is a typical hadronic scale, $L$ is the
linear size of the Euclidean box, and $m_\pi\sim\sqrt{m\Lambda}$ is
the pion mass).  The partition function was computed for equal quark
masses using an effective Lagrangian in Ref.~\cite{Leut92} and for the
general case of different quark masses in the framework of chiral RMT
in Ref.~\cite{Jack96c}.  The condition $L\ll 1/m_\pi$ sets an upper
limit of $1/\sqrt{V\Lambda^2}$ on the quark mass.  A lower limit of
$1/(V\Lambda^3)$ on the quark mass is set by the requirement that the
chiral condensate be non-zero in a finite volume.  To obtain
non-trivial results for the dependence of the various quantities we
compute on $m$, the quark mass has to be rescaled by the same factor
$1/(V\Sigma)$ as the eigenvalues, cf.~Eq.~(\ref{eq1.1}).  For large
quark masses, our results reduce to those obtained in the quenched
approximation.

This paper is organized as follows.  In Sec.~\ref{sec2} we discuss the
random-matrix model and construct the orthogonal polynomials and the
partition function which are needed in the course of the calculation.
Physical observables such as the microscopic spectral density and the
distribution of the smallest eigenvalue are computed in
Sec.~\ref{sec3}.  Section~\ref{sec4} is a short summary.

After completion of this work we learned that very recently a similar
calculation was performed simultaneously and independently by P.H.
Damgaard and S. Nishigaki \cite{Damg97}.  In this work, the authors
also give a universality proof for more general distributions of the
random matrix and construct spectral sum rules for the massive Dirac
operator (see also Ref.~\cite{Damg97b}) but do not address the
distribution of the smallest eigenvalue.  Wherever the two manuscripts
overlap the results are identical.

\section{Random-matrix theory for finite quark masses} 
\label{sec2}

In Sec.~\ref{sec21}, we formulate the chiral random-matrix model by
introducing the probability density function and the partition
function in the presence of finite quark masses. In Sec.~\ref{sec22},
we construct the orthogonal polynomials of the model. An alternative
construction is presented in Sec.~\ref{sec23} which is based on a certain
duality between models in different matrix spaces.

\subsection{Formulation of the model}
\label{sec21}

The statistical properties of the random-matrix model for the massive
Dirac operator with $N_f$ flavors and dynamical quark masses $M_f$
($f=1,\ldots,N_f$) which was introduced above are given by the
probability density function
\begin{equation}
P_N^{(N_f)}(W,M) \ = \ \frac{1}{\widetilde{Z}_N^{(N_f)}(M)} \, 
                \prod_{f=1}^{N_f}\det(WW^\dagger+M_f^2)
                \exp(-N\Sigma^2{{\rm tr}\,} WW^\dagger)
\label{pdf}
\end{equation}
with $M=(M_1,\ldots,M_{N_f}$). Here, $W$ is a complex matrix of
dimension $N$.  Apart from a Gaussian, this distribution contains
$N_f$ fermionic determinants. Eventually, we are mainly interested in
the microscopic limit. Hence, it suffices to take into account a
Gaussian in the total distribution (\ref{pdf}) since various authors
\cite{Brez96,Nish96} have given proofs of universality in related
problems.  Note that even for nonzero topological charge $\nu$ it is
sufficient to consider only square matrices $W$ in the calculation
because of the duality between topology and flavor
\cite{Verb93,Verb94a}.  A nonzero $\nu$ is obtained by introducing
$\nu$ additional massless flavors at the end of the calculation.  The
normalization constant
\begin{equation}
\widetilde{Z}_N^{(N_f)}(M) \ = \ 
               \int d[W] \prod_{f=1}^{N_f}\det(WW^\dagger+M_f^2)
                \exp(-N\Sigma^2{{\rm tr}\,} WW^\dagger)
\label{partf}
\end{equation}
depends on the quark masses and plays the role of a partition
function.  Since both of the above expressions depend on the matrices
$W$ only through invariant functions of $WW^\dagger$, they can be
reduced to integrals over radial coordinates. We write $W=U \Lambda
\bar{V}$ with $U\in{\rm U}(N)$, $\bar{V}\in{\rm U}(N)/ {\rm U}^N(1)$,
and $\Lambda={\rm diag}(\lambda_1,\ldots,\lambda_N)$, where the radial
coordinates $\lambda_i$ are restricted to the positive real axis.
Thus, we have $WW^\dagger=U\Lambda^2U^\dagger$ suggesting the
introduction of new variables $x_i=\lambda_i^2$. The measure
transforms as $d[W]=\Delta_N^2(X)d[X]d\mu(U)d\mu(\bar{V})$, where
$X={\rm diag} (x_1,\ldots,x_N)$, $\Delta_N(X)=\prod_{i<j}(x_i-x_j)$ is
the Vandermonde determinant, and $d\mu$ is the Haar measure.  After
integrating over the unitary groups, we obtain
\begin{equation}
\label{eq2.1}
P_N^{(N_f)}(X,M) = \frac{1}{Z_N^{(N_f)}(M)} \, \Delta_N^2(X) \, 
                 \prod_{i=1}^N w^{(N_f)}(x_i,M) 
\end{equation}
for the probability density and 
\begin{equation}
Z_N^{(N_f)}(M) = \int d[X] \, \Delta_N^2(X) \,
                 \prod_{i=1}^N w^{(N_f)}(x_i,M) 
\label{pdpfx}
\end{equation}
for the partition function $Z_N^{(N_f)}(M)$ which differs from
$\widetilde{Z}_N^{(N_f)}(M)$ by the group volumes. The expression
\begin{equation}
w^{(N_f)}(x,M) \ = \ \exp(-N\Sigma^2x) \prod_{f=1}^{N_f}(x+M_f^2) 
\label{weight}
\end{equation}
will be referred to as the weight function.

Because of a main result of RMT \cite{Meht91}, all spectral
correlation functions $R_k^{(N_f)}(x_1,\ldots,x_k,M)$ of our model can
readily be written in terms of the polynomials $p_n^{(N_f)}(x,M)$
which are orthogonal with respect to the weight function
$w^{(N_f)}(x,M)$,
\begin{equation}
  \int_0^\infty dx \, w^{(N_f)}(x,M) p_n^{(N_f)}(x,M) p_m^{(N_f)}(x,M)
  \ = \ 
  \frac{Z_{n+1}^{(N_f)}(M)}{(n+1)\,Z_n^{(N_f)}(M)} \, \delta_{nm} \ .
\label{ortho}
\end{equation} 
The correlation functions are given as the determinant
\begin{equation}
R_k^{(N_f)}(x_1,\ldots,x_k,M) \ = \ 
  \det\left[K_N^{(N_f)}(x_i,x_j,M)\right]_{i,j=1,\ldots,k}
\label{corrfct}
\end{equation}
with a kernel given by
\begin{equation}
K_N^{(N_f)}(x,y,M) \ = \ \sqrt{w^{(N_f)}(x,M)w^{(N_f)}(y,M)} \,
                 \sum_{n=0}^{N-1}   
                \frac{(n+1)\,Z_n^{(N_f)}(M)}{Z_{n+1}^{(N_f)}(M)}
                p_n^{(N_f)}(x,M) p_n^{(N_f)}(y,M) \ .
\label{summe}
\end{equation}
With the help of the Christoffel-Darboux formula the kernel
can be expressed as
\begin{eqnarray}
K_N^{(N_f)}(x,y,M) & = & N \, \frac{Z_{N-1}^{(N_f)}}{Z_N^{(N_f)}} \,
                 \sqrt{w^{(N_f)}(x,M)w^{(N_f)}(y,M)} \nonumber\\
  &&\times \frac{p_N^{(N_f)}(x,M)p_{N-1}^{(N_f)}(y,M)
    -p_{N-1}^{(N_f)}(x,M)p_N^{(N_f)}(y,M)}{x-y} \ ,
\label{kernel}
\end{eqnarray}
provided that the coefficient of the power $x^n$ in $p_n^{(N_f)}(x,M)$
is unity.  Eventually, we will take the microscopic limit as discussed
in the Introduction.

\subsection{Construction of orthogonal polynomials}
\label{sec22}

The polynomials $p_n^{(N_f)}(x,M)$ can be constructed by applying a
standard formula~\cite{Bate51} in the theory of orthogonal
polynomials. In our case, it reads
\begin{equation}
p_n^{(N_f)}(x,M) \ = \ \frac{1}{Z_n^{(N_f)}(M)} 
                     \int d[X] \Delta_n^2(X) 
                    \prod_{i=1}^n w^{(N_f)} (x_i,M) (x-x_i) \ .
\label{poly}
\end{equation}
This is an integral over $n$ variables $x_i$ ($i=1,\ldots,n$). We
notice that $n$ replaces the dimension $N$ of the matrices $W$ in the
partition function $Z_n^{(N_f)}(M)$ and in the Vandermonde determinant
$\Delta_n(X)$. Moreover, by construction, the coefficient in front
of $x^n$ is unity.

Fortunately, there is a further result, the Christoffel
formula~\cite{Bate51}, which allows us to immediately write down the
polynomials explicitly: The weight function $w^{(N_f)}(x,M)$ is a
product of the weight function $w^{(0)}(x)$ and a polynomial whose
zeros are the negative squared quark masses $-M_f^2$ and which is
nonnegative on the real axis.  Therefore, the polynomials
$p_n^{(N_f)}(x,M)$ can be written in terms of the polynomials which
are orthogonal with respect to the weight
$w^{(0)}(x)=\exp(-N\Sigma^2x)$.  These are precisely the Laguerre
polynomials $L_n^{(0)}(N\Sigma^2x)$.  The application of the
Christoffel formula then yields
\begin{eqnarray}
  p_n^{(N_f)}(x,M) &=& \frac{1}{C_n^{(N_f)}(M)} \,
  \frac{1}{\prod_{f=1}^{N_f}(x+M_f^2)} \,
  \frac{(n+N_f)!}{(N\Sigma^2)^{n+N_f}} \nonumber\\
& & \qquad \times\left| \begin{array}{llcl}
  L_n^{(0)}(N\Sigma^2x) & L_{n+1}^{(0)}(N\Sigma^2x) & \cdots &
                               L_{n+N_f}^{(0)}(N\Sigma^2x) \\ 
  L_n^{(0)}(-N\Sigma^2M_1^2) & L_{n+1}^{(0)}(-N\Sigma^2M_1^2) & \cdots &
                               L_{n+N_f}^{(0)}(-N\Sigma^2M_1^2) \\ 
     \qquad\vdots & \qquad\vdots & & \qquad\vdots \\ 
  L_n^{(0)}(-N\Sigma^2M_{N_f}^2) & L_{n+1}^{(0)}(-N\Sigma^2M_{N_f}^2)
   & \cdots &  L_{n+N_f}^{(0)}(-N\Sigma^2M_{N_f}^2) 
           \end{array}\right|  
\label{polcf}
\end{eqnarray} 
with a function $C_n^{(N_f)}(M)$ yet to be determined.  In
Sec.~\ref{sec23}, we will present a very direct and highly convenient
computation of the polynomials (\ref{polcf}) including all
normalization factors.  Here we proceed as follows: Since the
polynomial $p_n^{(N_f)}(x,M)$ has to be invariant under permutations
of the masses $M_f$ and must have a finite limit when all quark masses
are degenerate, a reasonable guess for the normalization function is
\begin{eqnarray}
C_n^{(N_f)}(M) \ = \ \det\left[L_{n+f-1}^{(0)}
        (-N\Sigma^2M_{f^\prime}^2)\right]_{f,f^\prime=1,\ldots,N_f}\:,
\label{polnorm}
\end{eqnarray} 
see Sec.~\ref{sec23} below, in particular Eq.~(\ref{chefformel}).
With this ansatz, the partition function $Z_n^{(N_f)}(M)$ is obtained
by studying the case of $N_f+1$ flavors. On the one hand,
Eqs.~(\ref{pdpfx}) and (\ref{poly}) yield
$p_n^{(N_f)}(-M_{N_f+1}^2,M)=-Z_n^{(N_f+1)}(M,M_{N_f+1})/Z_n^{(N_f)}(M)$,
which, on the other hand, has to be equal to the right-hand side of
Eq.~(\ref{polcf}) for $x=-M_{N_f+1}^2$. Thus, we arrive at the
recursion relation
\begin{equation}
\frac{Z_n^{(N_f+1)}(M,M_{N_f+1})}{C_n^{(N_f+1)}(M,M_{N_f+1})} 
\ = \ \frac{Z_n^{(N_f)}(M)}{C_n^{(N_f)}(M)} 
       \, \frac{(n+N_f)!}{(N\Sigma^2)^{n+N_f}} \,
          \frac{1}{\prod_{f=1}^{N_f}(M_f^2-M_{N_f+1}^2)} 
\label{recurs}
\end{equation} 
which implies
\begin{equation}
Z_n^{(N_f)}(M) \ = \ \prod_{f=1}^{N_f} \frac{(n+f-1)!}
{(N\Sigma^2)^{n+f-1}}\,\frac{C_n^{(N_f)}(M)}{\Delta_{N_f}(-M^2)}\:.
\label{parterg}
\end{equation} 
The correctness of these results can be verified using the
orthogonality relation~(\ref{ortho}).  In the microscopic limit, the
partition function agrees with the result obtained in
Ref.~\cite{Jack96c}, see Sec.~\ref{sec3}, Eq.~(\ref{CNfm}).  (Note
that in Ref.~\cite{Jack96c}, the result is expressed in terms of
derivatives of modified Bessel functions.  Using the recurrence
relations for Bessel functions \cite{Grad65} and the properties of the
determinant, one can show that the results are identical.)

\subsection{A duality of matrix ensembles}
\label{sec23}

In this section, we present an alternative derivation of the
orthogonal polynomials $p_n^{(N_f)}(x,M)$ and of the partition
function $Z_N^{(N_f)}(M)$. It employs a certain kind of duality
between matrix ensembles and is therefore of conceptual interest. On
the practical side, this method advantageously yields all
normalizations, particularly the function $C_n^{(N_f)}(M)$ and the
partition function $Z_n^{(N_f)}(M)$, in a very direct and convenient
way. We consider the integral
\begin{equation}
X_N^{(K)}(a) \ = \int d[W] \, \exp(-\alpha{{\rm tr}\,} WW^\dagger) \,
\prod_{f=1}^K \det\left[
            \begin{array}{cc} -ia_f & W \\ 
                          W^\dagger & -ia_f 
            \end{array} \right]
\label{Xdef}
\end{equation}
depending on $K$ parameters $a_f$ ($f=1,\ldots,K$). For
$\alpha=N\Sigma^2$, $K=N_f$, and $a_f=M_f$, the integral~(\ref{Xdef})
is just the partition function~(\ref{partf}). For $N=n$, $K=N_f+1$,
$a_f=M_f$ ($f=1,\ldots,N_f$), and $a_{N_f+1}^2=-x$, the
integral~(\ref{Xdef}) is, up to the group volumes, equal to
$Z_n^{(N_f)}(M)p_n^{(N_f)}(x,M)$ as defined in Eq.~(\ref{poly}). Thus,
the function $X_N^{(K)}(a)$ yields all desired quantities.  We define
$K$ $2N$-component vectors $\zeta_f$ ($f=1,\ldots,K$) with complex
Grassmannian entries $\zeta_{fi}$ ($i=1,\ldots,2N$) as well as the
$2NK$-component combination $\zeta=(\zeta_1,\ldots,\zeta_K)^T$.  The
determinants in the integrand of Eq.~(\ref{Xdef}) are represented as
the integral
\begin{equation}
\prod_{f=1}^K \det\left[\begin{array}{cc} -ia_f      & W \\ 
                                           W^\dagger & -ia_f 
                        \end{array}\right] \ = \ 
(2\pi)^{2NK} \, \int d[\zeta] \, \exp\left(\zeta^\dagger
  \left(\openone_K\otimes
   \left[\begin{array}{cc} 0         & W \\ 
                           W^\dagger & 0
    \end{array}\right] - ia\otimes \openone_{2N}\right)\zeta\right)
\label{fermint}
\end{equation}
with $a={\rm diag}(a_1,\ldots,a_K)$. In this form, the Gaussian
average over the matrices $W$ can easily be performed. We arrive at a
four-fermion interaction which is then decoupled by a
Hubbard-Stratonovitch transformation. In Ref.~\cite{Guhr97a}, a
very similar route was taken. In this study, however, a problem
involving fermionic and bosonic determinants was solved which led to
the introduction of supervectors of commuting and anticommuting
variables. In the present case, the vectors $\zeta_f$ have only
anticommuting entries. Fortunately, this difference has little
influence on the structure of the results. Instead of the
supermatrices which were used in Ref.~\cite{Guhr97a} for the
Hubbard-Stratonovitch transformation, we have to employ ordinary
matrices here. We arrive at
\begin{equation}
X_N^{(K)}(a) \ = \ \left(\frac{\alpha}{\pi}\right)^{K^2}
                   \, \int d[\sigma]
  \exp\left(-\alpha{{\rm tr}\,}(\sigma-a)(\sigma^\dagger-a)\right)
                   \, {\rm det}^N\sigma\sigma^\dagger \:,
\label{dualmod}
\end{equation}
where $\sigma$ is an ordinary complex $K\times K$ matrix without
further symmetries. Thus, we have mapped the original model of
Sec.~\ref{sec21} in the form~(\ref{Xdef}) in the space of the $N\times
N$ matrices $W$ onto a dual model in the space of the $K\times K$
matrices $\sigma$. This is highly convenient because, eventually, we
wish to take the limit $N\to\infty$.  Remarkably, these dual models
differ with respect to rotation invariance. The model of
Sec.~\ref{sec21} is invariant under rotations in the sense that the
integrand of the partition functions in Eqs.~(\ref{partf})
and~(\ref{pdpfx}) depends only on the radial coordinates of the
matrices $W$, i.e., on the variables $x_i$.  This is not true for the
model~(\ref{dualmod}). The presence of the matrices $a$ breaks this
kind of rotation invariance. In Refs.~\cite{Guhr96b} and
\cite{Guhr97a}, a procedure was developed to deal with models
precisely of this kind. We introduce radial coordinates by
$\sigma=us\bar{v}$ with $u\in {\rm U}(K)$, $\bar{v}\in {\rm U}(K)/
{\rm U}^K(1)$, and $s={\rm diag}(s_1,\ldots,s_K)$. Again, the radial
coordinates $s_f$ are restricted to the positive real axis. The
measure transforms as
$d[\sigma]=\Delta_K^2(s^2)\prod_{f=1}^Ks_fd[s]d\mu(u)d\mu(\bar{v})$.
The integral over the unitary groups is now nontrivial. The solution
can be found in Ref.~\cite{Guhr96b},
\begin{equation}
\left(\frac{\alpha}{\pi}\right)^{K^2} \, 
          \int d\mu(u) \int d\mu(\bar{v}) 
          \exp\left(-\alpha{{\rm tr}\,}(us\bar{v}-a)
                   (\bar{v}^\dagger su^\dagger-a)\right) \ = \
           \frac{1}{K!}
           \frac{\det\left[\gamma(s_f,a_{f^\prime})
                      \right]_{f,f^\prime=1,\ldots,K}}
                {\Delta_K(s^2)\Delta_K(a^2)} \ .
\label{groupint}
\end{equation}
The entries of the determinant are given by
\begin{equation}
\gamma(s_f,a_{f^\prime}) \ = \ 2\alpha \, 
         \exp\left(-\alpha(s_f^2+a_{f^\prime}^2)\right) \,
                        I_0(2\alpha s_f a_{f^\prime}) \ ,
\label{diffkern}
\end{equation}
where $I_0$ is the modified Bessel function of the first kind and
zeroth order.  Collecting everything, we obtain
\begin{equation}
X_N^{(K)}(a) \ = \ \frac{1}{K!} \, \frac{1}{\Delta_K(a^2)} 
       \prod_{f=1}^K \int_0^\infty ds_f \, s_f^{2N+1} \, \Delta_K(s^2)
  \det\left[\gamma(s_f,a_{f^\prime})\right]_{f,f^\prime=1,\ldots,K}\:.
\label{integral}
\end{equation}
This integral can be solved in a straightforward way. We write
$\Delta_K(s^2)=\det[s_f^{2(f^\prime-1)}]_{f,f^\prime=1,\ldots,K}$
and integrate this determinant row by row. Using the representation
\begin{equation}
2\alpha \int_0^\infty dw \, w^{2n+2m+1} \,
       \exp\left(-\alpha(z^2+w^2)\right) I_0(2\alpha wz) \ = \
\frac{(n+m)!}{\alpha^{n+m}} \, L_{n+m}^{(0)}(-\alpha z^2)
\label{intrepres}
\end{equation}
for the Laguerre polynomials, we find 
\begin{equation}
X_N^{(K)}(a) \ = \ \frac{1}{\Delta_K(a^2)} 
 \prod_{f=0}^{K-1} \frac{(N+f)!}{\alpha^{N+f}}
 \det\left[L_{N+f-1}^{(0)}(-\alpha
 a_{f'}^2)\right]_{f,f^\prime=1,\ldots,K} \ .
\label{chefformel}
\end{equation}
This result immediately yields the orthogonal polynomials
$p_n^{(N_f)}(x,M)$ and the partition function $Z_N^{(N_f)}(M)$.

\section{Calculation of observables}
\label{sec3}

The microscopic spectral correlators, in particular the microscopic
spectral density, are computed in Sec.~\ref{sec31}. In
Sec.~\ref{sec32}, we derive the distribution of the smallest
eigenvalue.

\subsection{Microscopic spectral correlators and density}
\label{sec31}

To calculate the spectral correlations, we insert the polynomials
$p_n^{(N_f)}(x,M)$ in Eq.~(\ref{kernel}) for the kernel
$K_N^{(N_f)}(x,y,M)$.  The microscopic limit is obtained according to
Eq.~(\ref{eq1.1}) by rescaling energies and quark masses by
$2N\Sigma$.  We define $\lambda=2N\Sigma \sqrt{x}$ and $m_f=2N\Sigma
M_f$.  In the limit $N\to\infty$, we want both $\lambda$ and $m_f$ to
be of order unity.  The quenched approximation then corresponds to
taking $m_f\to\infty$ (or, of course, setting $N_f=0$).  In the limit
$m_f\to 0$, our results must reduce to those obtained earlier in the
chiral limit \cite{Verb93}.  After rescaling the parameters one uses
the asymptotic relation $\lim_{N\to\infty}L_N^{(\alpha)}(z/N)/N^\alpha
= z^{-\alpha/2}J_\alpha(2\sqrt{z})$, where $J_\alpha$ denotes the
Bessel function of the first kind and order $\alpha$. To obtain the
correct large-$N$ limit one has to use the recurrence relations for
the Laguerre polynomials \cite{Grad65} in Eq.~(\ref{polcf}).  The
properties of the determinant can be used to simplify the expression.
After summing up one uses
\begin{eqnarray}
{\left(\frac{z^2}{N}\right)^\alpha}
        L_{N-1}^{(\alpha)}{\left(\frac{z^2}{N}\right)}&=&
{\frac{N}{N+\alpha}\left(\frac{z^2}{N}\right)^\alpha}
        L_N^{(\alpha)}{\left(\frac{z^2}{N}\right)}\nonumber\\
&&+{\frac{1}{N+\alpha}\left(\frac{z^2}{N}\right)^{\alpha+1}}
        L_N^{(\alpha+1)}{\left(\frac{z^2}{N}\right)}
-{\frac{1}{N+\alpha}\left(\frac{z^2}{N}\right)^{\alpha+1}}
        L_N^{(\alpha)}{\left(\frac{z^2}{N}\right)}\:.
\label{reccur}
\end{eqnarray}
Note that the first, second, and third term on the right-hand side is
of ${\cal O}(1)$, ${\cal O}(1/N)$, and ${\cal O}(1/N^2)$,
respectively.  Using the properties of the determinant and collecting
the terms of  ${\cal O}(1)$ carefully, one obtains the
microscopic limit of (\ref{kernel}).  We arrive at
\begin{eqnarray}
K^{(N_f)}(\lambda,\lambda^\prime,m) & = &\lim\limits_{N\to\infty}
 \:\frac{2\sqrt{\lambda\lambda^\prime}}{(2N\Sigma)^2}\:
  K_N^{(N_f)}\left(\left(\lambda/2N\Sigma\right)^2,
    \left(\lambda^\prime/2N\Sigma\right)^2,
    m/2N\Sigma\right)
  \nonumber \\ & & \hspace*{-35pt} = 
  \frac{\sqrt{\lambda\lambda^\prime}}{\lambda^{\prime 2}-\lambda^2}
  \frac{B^{(N_f)}(\lambda,m)\sum_{f=0}^{N_f}
  \widetilde{B}_f^{(N_f)}(\lambda^\prime,m)
  -B^{(N_f)}(\lambda^\prime,m)\sum_{f=0}^{N_f}
  \widetilde{B}_f^{(N_f)}(\lambda,m)}
  {[C^{(N_f)}(m)]^2\prod_{f=1}^{N_f}\sqrt{(\lambda^2+m_f^2)
      (\lambda^{\prime 2}+m_f^2)}}\:,
\label{microkernel}
\end{eqnarray}
where the normalization $C^{(N_f)}(m)$ is the microscopic limit
of the normalization $C_N^{(N_f)}(M)$ of Eq.~(\ref{polnorm}),
\begin{equation}
C^{(N_f)}(m) \ = \ \lim_{N\to\infty} C_N^{(N_f)}(m/2N\Sigma)
             \ = \ \det\left[m_{f^\prime}^{f-1}I_{f-1}(-m_{f^\prime})
                                 \right]_{f,f^\prime=1,\ldots,N_f} \ .
\label{CNfm}
\end{equation}
The determinant $B^{(N_f)}(\lambda,m)$ in Eq.~(\ref{microkernel}) is
given by
\begin{equation}
B^{(N_f)}(\lambda,m)=\left|
\begin{array}{cccc}
J_0(\lambda) & I_0(-m_1) & \cdots & I_0(-m_{N_f}) \\
\lambda J_1(\lambda) & m_1I_1(-m_1) & \cdots & m_{N_f}I_1(-m_{N_f}) \\
\vdots & \vdots & & \vdots \\
\lambda^{N_f}J_{N_f}(\lambda) & m_1^{N_f}I_{N_f}(-m_1) & \cdots &
m_{N_f}^{N_f}I_{N_f}(-m_{N_f}) \end{array} \right| \ .
\label{matrixb}
\end{equation}
The determinant $\widetilde{B}_f^{(N_f)}(\lambda,m)$ is constructed
from $B^{(N_f)}(\lambda,m)$ by replacing the first column by 
$(\lambda J_1(\lambda), \ldots,\lambda^{N_f+1}J_{N_f+1}(\lambda))^T$ 
if $f=0$ and by replacing the ($f$+1)-th column by
$(m_fI_1(-m_f),\ldots,m_f^{N_f+1}I_{N_f+1}(-m_f))^T$ if
$f=1,\ldots,N_f$.  All $k$-point spectral correlations can be obtained
from the kernel (\ref{microkernel}) in analogy with
Eq.~(\ref{corrfct}).  Most importantly, the microscopic spectral
density is obtained by taking the limit $\lambda'=\lambda=z$ in
(\ref{microkernel}).  This yields
\begin{equation}
  \label{microdens}
  \rho_s^{(N_f)}(z,m)=\frac{z}{2}
  \frac{B^{(N_f)}(z,m)\sum_{f=0}^{N_f}\widetilde{D}_f^{(N_f)}(z,m)
  -D^{(N_f)}(z,m)\sum_{f=0}^{N_f}\widetilde{B}_f^{(N_f)}(z,m)}
  {[C^{(N_f)}(m)]^2\prod_{f=1}^{N_f}(z^2+m_f^2)}\:,
\end{equation}
where the determinant $D^{(N_f)}(z,m)$ is constructed from
$B^{(N_f)}(z,m)$ by replacing the first column in (\ref{matrixb}) by
$(z^{-1}J_{-1}(z),\ldots, z^{N_f-1}J_{N_f-1}(z))^T$,
$\widetilde{D}_f^{(N_f)}(z,m)$ is constructed from $D^{(N_f)}(z,m)$ by
replacing the ($f$+1)-th column by
$(m_fI_1(-m_f),\ldots,m_f^{N_f+1}I_{N_f+1}(-m_f))^T$ if
$f=1,\ldots,N_f$, and $\widetilde{D}_0^{(N_f)}(z,m)=B^{(N_f)}(z,m)$.

\begin{figure}
  \centerline{\psfig{figure=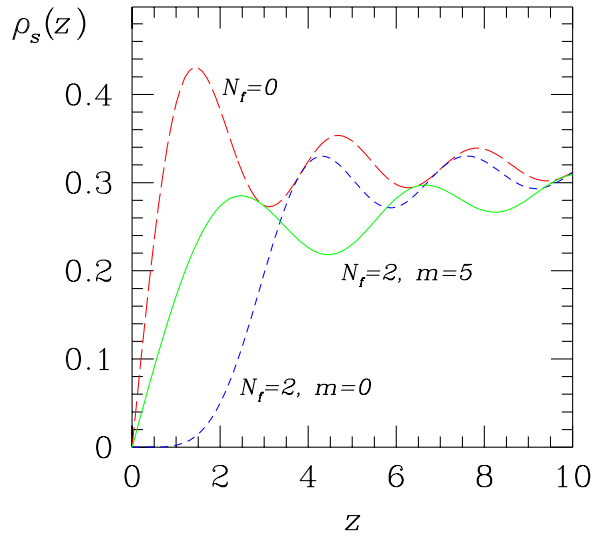,width=80mm}}
  \caption{Microscopic spectral density for two
    degenerate flavors of rescaled mass $m=5$ (solid line), in the
    chiral limit (short dashes), and in the quenched
    approximation (long dashes).}
  \label{fig1}
\end{figure}

We have verified that our results (\ref{microkernel}) and
(\ref{microdens}) reproduce the known results for the chiral limit and
the quenched approximation if $m_f\to 0$ and $m_f\to\infty$,
respectively.  In Refs.~\cite{Guhr97a,Jack97a}, the correlations of
the Dirac operator at finite temperature were studied using the
supersymmetry method. These investigations suggest that the
expression~(\ref{microkernel}) for the kernel can be simplified
further and expressed in form of a single determinant.
Indeed, such a more compact form was very recently obtained in
Ref.~\cite{Damg97}.  We write the result in the form
\begin{eqnarray}
K^{(N_f)}(\lambda,\lambda^\prime,m) & = &  
  \frac{\sqrt{\lambda\lambda^\prime}}{\lambda^{\prime 2}-\lambda^2}
  \frac{1}{C^{(N_f)}(m)\prod_{f=1}^{N_f}\sqrt{(\lambda^2+m_f^2)
      (\lambda^{\prime 2}+m_f^2)}}
  \nonumber \\ & & \hspace*{-60pt}\times
\left|\begin{array}{ccccc}
J_0(\lambda) & J_0(\lambda^\prime) & I_0(-m_1) & \ldots & I_0(-m_{N_f}) \\
\lambda J_1(\lambda) & \lambda^\prime J_1(\lambda^\prime) & 
  m_1I_1(-m_1) & \ldots & m_{N_f}I_1(-m_{N_f}) \\
\vdots & \vdots & \vdots & & \vdots \\
\lambda^{N_f+1} J_{N_f+1}(\lambda) & 
  {\lambda^\prime}^{N_f+1} J_{N_f+1}(\lambda^\prime) & 
  m_1^{N_f+1}I_{N_f+1}(-m_1) & \ldots & 
  m_{N_f}^{N_f+1}I_{N_f+1}(-m_{N_f}) 
\end{array}\right| \ .
\label{darmnij}
\end{eqnarray}
The microscopic spectral density then reads
\begin{eqnarray}
\rho_s^{(N_f)}(z,m) & = &  
  -\frac{z}{2}
  \frac{1}{C^{(N_f)}(m)\prod_{f=1}^{N_f}(z^2+m_f^2)}\nonumber \\ 
  & & \hspace*{-22pt}\times
\left|\begin{array}{ccccc}
  z^{-1}J_{-1}(z) & J_0(z) & I_0(-m_1) & \ldots & I_0(-m_{N_f}) \\
  J_0(z) & z J_1(z) & 
  m_1I_1(-m_1) & \ldots & m_{N_f}I_1(-m_{N_f}) \\
  \vdots & \vdots & \vdots & & \vdots \\
  z^{N_f} J_{N_f}(z) & 
  z^{N_f+1} J_{N_f+1}(z) & 
  m_1^{N_f+1}I_{N_f+1}(-m_1) & \ldots & 
  m_{N_f}^{N_f+1}I_{N_f+1}(-m_{N_f}) 
\end{array}\right|\:.
\label{mdDN}
\end{eqnarray}
We have verified by direct calculation that the expressions
(\ref{microkernel}), (\ref{darmnij}) and (\ref{microdens}),
(\ref{mdDN}) are identical, respectively.  A number of interesting
special cases can be derived from (\ref{microdens}) or (\ref{mdDN}).
(For $N_f=1$ and $\nu=0$, $\rho_s$ was first computed in
Ref.~\cite{Jurk96}.  However, like the authors of \cite{Damg97}, we
fail to agree with the result given in Eq.~(80) of Ref.~\cite{Jurk96}
which does not reproduce the quenched result in the limit
$m\to\infty$.)  As an example, we give the result for $\rho_s$ for two
degenerate flavors of mass $m$ and $\nu=0$,
\begin{equation}
  \rho_s^{(2)}(z,m)=\frac{z}{2}\left(J_0^2(z)+J_1^2(z)\right)
  -\frac{2z}{(z^2+m^2)^2}\frac{[zJ_1(z)I_0(m)+mJ_0(z)I_1(m)]^2}
  {I_0^2(m)-I_1^2(m)}\:.
\end{equation}
This expression is plotted in Fig.~\ref{fig1} compared with the
results obtained in the chiral limit and in the quenched
approximation.  
Clearly, the presence of light flavors leads to a suppression of small
eigenvalues as expected.  We emphasize again that while we have worked
with $\nu=0$, the general case $\nu>0$ can easily be obtained by
introducing $\nu$ additional massless flavors. For completeness, we
state the result for the microscopic spectral density for general
$\nu$ which is obtained by expanding the determinants in (\ref{mdDN}),
\begin{eqnarray}
\rho_s^{(N_f,\nu)}(z,m) & = &  
  -\frac{z}{2}
  \frac{1}{\widetilde{C}^{(N_f,\nu)}(m)z^{2\nu}
    \prod_{f=1}^{N_f}(z^2+m_f^2)}  \nonumber \\ 
  & & \hspace*{-80pt} \times
\left|\begin{array}{ccccc}
  z^{\nu-1}J_{\nu-1}(z) & z^\nu J_\nu(z) & m_1^\nu I_\nu(-m_1) & 
  \ldots & m_{N_f}^\nu I_\nu(-m_{N_f}) \\
  z^\nu J_\nu(z) & z^{\nu+1} J_{\nu+1}(z) & 
  m_1^{\nu+1}I_{\nu+1}(-m_1) & \ldots & 
  m_{N_f}^{\nu+1}I_{\nu+1}(-m_{N_f}) \\
  \vdots & \vdots & \vdots & & \vdots \\
  z^{N_f+\nu} J_{N_f+\nu}(z) & 
  z^{N_f+\nu+1} J_{N_f+\nu+1}(z) & 
  m_1^{N_f+\nu+1}I_{N_f+\nu+1}(-m_1) & \ldots & 
  m_{N_f}^{N_f+\nu+1}I_{N_f+\nu+1}(-m_{N_f}) 
\end{array}\right|
\label{rhosnu}
\end{eqnarray}
with
\begin{equation}
  \widetilde{C}^{(N_f,\nu)}(m)=
  \det\left[m_{f^\prime}^{\nu+f-1}I_{\nu+f-1}(-m_{f^\prime}) 
        \right]_{f,f^\prime=1,\ldots,N_f}\:.
\end{equation}
Note that the chiral limit follows immediately from (\ref{rhosnu}) by
the replacements $N_f=0$ and $\nu\to N_f+\nu$.   We then obtain
\begin{equation}
  \rho_s^{(N_f,\nu)}(z,0)=\frac{z}{2}\left(J_{N_f+\nu}^2(z)
    -J_{N_f+\nu+1}(z)J_{N_f+\nu-1}(z)\right)
\end{equation}
as required \cite{Verb93}.

\subsection{Distribution of the smallest eigenvalue}
\label{sec32}

In lattice simulations, the performance of an algorithm is frequently
determined by the magnitude of the smallest eigenvalue.  It is
therefore of interest to compute an analytical result for the
distribution of the smallest eigenvalue, $P^{(N_f)}(\lambda_{\rm
  min},m)$, in the framework of RMT.  This is the subject of this
section.  The RMT result should be universal and identical with that
of full QCD under the same conditions as the microscopic spectral
density.  The only result that has been computed so far in the
literature applies to the quenched approximation where
$P^{(0)}(\lambda)=\lambda\exp(-\lambda^2/4)/2$ \cite{Forr93}.

\begin{figure}[ht]
  \centerline{\psfig{figure=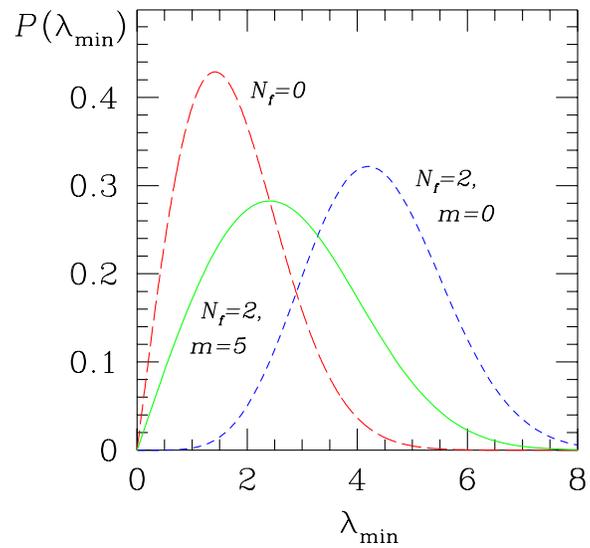,width=80mm}}
  \caption{Distribution of the smallest eigenvalue for two
    degenerate flavors of rescaled mass $m=5$ (solid line), in the
    chiral limit (short dashes), and in the quenched
    approximation (long dashes).}
  \label{fig2}
\end{figure}

A well-known quantity in RMT is the so-called ``hole probability''
$E(s_1,s_2)$ which is the probability that the interval $(s_1,s_2)$ is
free of eigenvalues.  For the smallest eigenvalue, we again move to
the microscopic scale by rescaling energies and masses by $2N\Sigma$.
For the interval $(0,s)$ at the spectrum edge, we then have
\begin{equation}
  \label{eq4.1}
  E(0,s)=\int_s^\infty dx_1\cdots dx_N \widetilde\rho_N^{(N_f)}
  (x_1,\ldots,x_N;m)\:,
\end{equation}
where the probability density function on the microscopic scale (with
$x_i=z_i^2$) is given by
\begin{equation}
  \label{eq4.2}
  \widetilde\rho_N^{(N_f)}(x_1,\ldots,x_N;m)=c_N^{(N_f)}(m)
  \Delta_N^2(x)\prod_{i=1}^Ne^{-x_i/4N} \prod_{f=1}^{N_f}(x_i+m_f^2)\:,
\end{equation}
cf.~Eqs.~(\ref{eq2.1}) and (\ref{weight}).  The factor
$c_N^{(N_f)}(m)$ ensures that $E(0,0)=1$ and should not be confused
with the normalization constant defined in Eqs.~(\ref{polnorm}) and
(\ref{CNfm}).  On the other hand, we have
\begin{equation}
  \label{eq4.3}
  E(0,s)=\int_s^\infty dx_{\rm min} P^{(N_f)}(x_{\rm min},m)\:.
\end{equation}
Hence, the distribution of the smallest eigenvalue ($x_{\rm min}
=\lambda_{\rm min}^2$) is given by
\begin{equation}
  \label{eq4.4}
  P^{(N_f)}(\lambda_{\rm min},m)=2\lambda_{\rm min}
  P^{(N_f)}(x_{\rm min},m)=
  -2\lambda_{\rm min}\left.\frac{dE(0,s)}{ds}
  \right|_{s=\lambda_{\rm min}^2} \:.
\end{equation}
Performing the derivatives, we obtain
\begin{eqnarray}
  \label{eq4.5}
  -E'(0,s)&=&c_N^{(N_f)}(m)Ne^{-s/4N}\prod_{f=1}^{N_f}(s+m_f^2)
  \nonumber\\ 
  &&\times\int_s^\infty dx_1\cdots dx_{N-1} \Delta_{N-1}^2(x)
  \prod_{i=1}^{N-1}
  e^{-x_i/4N} (x_i-s)^2 \prod_{f=1}^{N_f}(x_i+m_f^2) \:.
\end{eqnarray}
We now shift $x_i\to x_i+s$ and finally obtain from (\ref{eq4.4})
\begin{eqnarray}
  \label{eq4.7}
  P^{(N_f)}(\lambda,m)&=&2c_N^{(N_f)}(m)N\lambda e^{-\lambda^2/4}
  \prod_{f=1}^{N_f}(\lambda^2+m_f^2)\nonumber\\
  &&\times\int_0^\infty dx_1\cdots dx_{N-1} \Delta_{N-1}^2(x) 
  \prod_{i=1}^{N-1}
  e^{-x_i/4N} x_i^2 \prod_{f=1}^{N_f}(x_i+\lambda^2+m_f^2)\:,
\end{eqnarray}
where we have written $\lambda$ instead of $\lambda_{\rm min}$ for
brevity.  The integrals in Eq.~(\ref{eq4.7}) can be done by comparison
with the case of the microscopic spectral density.  In terms of the
probability density function on the microscopic scale, $\rho_s$ is
given by
\begin{eqnarray}
  \rho_s^{(N_f)}(z,m)&=& 2z\cdot N\int_0^\infty dx_1\cdots dx_{N-1} 
  \widetilde\rho_N^{(N_f)}(x_1,\ldots,x_{N-1},z^2;m)\nonumber\\
  &=&2c_N^{(N_f)}(m)Nz\prod_{f=1}^{N_f}(z^2+m_f^2) \nonumber\\
  &&\times\int_0^\infty dx_1\cdots dx_{N-1}\Delta_{N-1}^2(x)
  \prod_{i=1}^{N-1}
  e^{-x_i/4N} (x_i-z^2)^2 \prod_{f=1}^{N_f}(x_i+m_f^2)
  \label{eq4.8}
\end{eqnarray}
in the large-$N$ limit.  Apart from prefactors in front of the
integrals, we note that $P^{(N_f)}(\lambda,m)$ can be obtained by
setting $z=0$ in the integrand of Eq.~(\ref{eq4.8}) and by replacing
$m_f$ by $(\lambda^2+m_f^2)^{1/2}$ in the result for $\rho_s$ (but not
in the normalization factors).  Setting $z=0$ in the numerator
determinant of Eq.~(\ref{mdDN}) makes the first two columns trivial,
and the dimension of the matrix can be reduced to $N_f$ by expanding
the determinant.  Including the prefactors and the proper
normalization, we obtain
\begin{equation}
  \label{eq4.9}
  P^{(N_f)}(\lambda,m)=\frac{\lambda}{2}e^{-\lambda^2/4}
  \frac{\det\left[\left(\lambda^2+m_{f^\prime}^2\right)^{(f+1)/2}
        I_{f+1}\left(-(\lambda^2+m_{f^\prime}^2)^{1/2}\right)
        \right]_{f,f^\prime=1,\ldots,N_f}}
  {C^{(N_f)}(m)\prod_{f=1}^{N_f}(\lambda^2+m_f^2)}
\end{equation}
with $C^{(N_f)}(m)$ given in Eq.~(\ref{CNfm}).  For $N_f=0$,
Eq.~(\ref{eq4.9}) reduces to the quenched result trivially.  We also
reproduce the quenched result by taking the limit $m_f\to\infty$.  The
simplest nontrivial case is $N_f=1, \nu=0$ where
\begin{equation}
  \label{eq4.10}
  P^{(1)}(\lambda,m)=\frac{\lambda}{2}e^{-\lambda^2/4}
  \frac{I_2\left(\sqrt{\lambda^2+m^2}\right)}{I_0(m)}\:.
\end{equation}
For two degenerate flavors of mass $m$ and $\nu=0$, we obtain
\begin{equation}
  \label{eq4.11}
  P^{(2)}(\lambda,m)=
  \frac{\lambda}{2}e^{-\lambda^2/4}
  \frac{1}{I_0^2(m)-I_1^2(m)}
  \left[I_2^2(t)-I_1(t)I_3(t)\right]_{t=\sqrt{\lambda^2+m^2}}\:.
\end{equation}
In Fig.~\ref{fig2}, we plot the distribution of the smallest
eigenvalue for the same parameters as in Fig.~\ref{fig1}. We observe
the same suppression of small eigenvalues in the presence of light
dynamical quarks.
Other cases can readily be derived from (\ref{eq4.9}).  The most
general case where $\nu>0$ can again be obtained by introducing $\nu$
additional massless flavors and expanding the determinants in
(\ref{eq4.9}).  As an example, we state the result for one massive
flavor and $\nu=1$,
\begin{equation}
  P^{(N_f=1,\nu=1)}(\lambda,m)=\frac{\lambda}{2}
  e^{-\lambda^2/4}\frac{1}{mI_1(m)}
  \left[tI_2(\lambda)I_3(t)-\lambda I_2(t)I_3(\lambda)
  \right]_{t=\sqrt{\lambda^2+m^2}}\:.
\end{equation}

\section{Discussion}
\label{sec4}

We have computed the microscopic spectrum of the massive QCD Dirac
operator, including the microscopic spectral density and the
distribution of the smallest eigenvalue, for an arbitrary number of
flavors, arbitrary quark masses, and arbitrary topological charge.
Our results generalize the previously known results for the
microscopic correlations in the chiral limit \cite{Verb93}.  The
distribution of the smallest eigenvalue was previously known only in
the quenched approximation.  Wherever there is an overlap, our results
agree with those obtained in the recent preprint by Damgaard and
Nishigaki \cite{Damg97}.  It would be very interesting to obtain the
result (\ref{darmnij}) using the supersymmetry method as in
Refs.~\cite{Guhr97a,Jack97a}.  Work in this direction is in progress.

The quantities we have computed in this paper are conjectured to be
universal, i.e., identical to those of QCD, for quark masses of the
order of $1/V$.  This conjecture can be verified straightforwardly in
lattice calculations.  One computes the eigenvalues of the Dirac
matrix and constructs the microscopic spectral density and the
distribution of the smallest eigenvalue by averaging over a sufficient
number of independent gauge field configurations.  The lattice data
can then be compared to the analytical predictions of RMT.  For pure
SU(2) gauge theory without dynamical quarks, this was already done in
Ref.~\cite{Berb97a}, and the agreement was excellent.

Our present calculation applies to full QCD with 3 colors in the
continuum limit.  In the context of lattice QCD, it applies to SU(3)
with dynamical staggered fermions.  The necessary lattice algorithms
are available, e.g., the hybrid Monte-Carlo algorithm for generating
unquenched configurations and the Lanczos algorithm for computing
eigenvalues of large sparse matrices.  Therefore, we are confident
that the results obtained in this paper will be verified in the very
near future.

Given that the analytical results obtained in this paper really
describe QCD, one may ask what their relevance is in practice.  We
list here a few possible applications.  (i) The parameter which
appears in the microscopic quantities contains the absolute value of
the chiral condensate, $\Sigma$.  As we have already seen in quenched
SU(2) \cite{Berb97b}, the analytical information available from RMT
can be used to determine $\Sigma$ and to extract its thermodynamic
limit from very small lattices.  Since otherwise one would have to
perform an extrapolation, this seems to offer an interesting
technological advantage.  (ii) Lattice simulations with light
dynamical quarks are very expensive, and extrapolations to the chiral
limit are difficult.  In RMT, we have computed exact analytical
expressions which apply for very small quark masses, and the chiral
limit can be taken analytically.  One would, therefore, hope that
lattice practitioners can use the analytical information in their
extrapolations to the chiral limit for those quantities that depend on
the small eigenvalues.  (iii) A very interesting issue is topology.
As one approaches the continuum limit on the lattice, one expects zero
modes to appear which give rise to non-zero topological charge $\nu$.
The microscopic quantities we consider are sensitive to topology, and
we have computed them for arbitrary $\nu$.  Hence, one can analyze
lattice results in different sectors of topological charge.  Whether
or not topology is seen on present-day lattices is still a somewhat
controversial issue, and for quenched SU(2) the lattice results of
Ref.~\cite{Berb97a} were consistent with $\nu=0$.  We hope that our
results will prove useful in future analyses of topology.  (iv) The
performance of many lattice algorithms depends on the magnitude of the
smallest eigenvalue.  While RMT cannot predict this magnitude, it
predicts the distribution of $\lambda_{\rm min}$ over the ensemble of
gauge field configurations.  It would be interesting to see if experts
in algorithm development can make use of the analytical information
available from RMT.

\section*{Acknowledgments}

It is a pleasure to thank P.H. Damgaard, J.J.M. Verbaarschot, and H.A.
Weidenm\"uller for stimulating discussions.  T. Wettig acknowledges
the hospitality of the MPI Heidelberg.  This work was supported in
part by DFG grant We 655/11-2.

\end{document}